\documentclass[prc,showpacs,superscriptaddress,twocolumn]{revtex4}

\usepackage{graphicx}
\usepackage{epstopdf}
\usepackage{subfigure}
\usepackage{epsfig}
\usepackage{hyperref}
\usepackage{amsmath,amssymb,amsfonts,bm}
\usepackage{color}
\usepackage[utf8]{inputenc}
\usepackage{verbatim}
\usepackage{array}
\usepackage{float}
\usepackage{lipsum}

\begin{document}
\newcommand{\dd}{\textrm{\,d}}

\newcommand{\pp}{p\kern-0.05em p}
\newcommand{\ppbar}{\mathrm{p}\kern-0.05em \overline{\mathrm{p}}}
\newcommand{\pPb}{\ensuremath{\mbox{p--Pb}}}
\newcommand{\PbPb}{\ensuremath{\mbox{Pb--Pb}}}
\newcommand{\GeV}{\ensuremath{\mathrm{GeV}\kern-0.05em}}
\newcommand{\GeVc}{\ensuremath{\mathrm{GeV}\kern-0.05em/\kern-0.02em c}}
\newcommand{\sqrts}{\ensuremath{\sqrt{s_{\mathrm{NN}}}}}
\newcommand{\pT}{\ensuremath{p_{\mathrm{T}}}}
\newcommand{\RL}{\ensuremath{R_{\mathrm{L}}}}
\newcommand{\pTi}{\ensuremath{p_{\mathrm{T},i}}}
\newcommand{\pTtrack}{\ensuremath{p_{\mathrm{T,track}}}}
\newcommand{\sigmaeec}{\ensuremath{\Sigma_{\mathrm{EEC}}}}

\newcommand{\kT}{\ensuremath{k_{\mathrm{T}}}}
\newcommand{\pThard}{\ensuremath{p_{\mathrm{T,hard}}}}
\newcommand{\etajet}{\ensuremath{\eta_{\mathrm{jet}}}}
\newcommand{\pTjet}{\ensuremath{p_{\mathrm{T}}^{\mathrm{jet}}}}
\newcommand{\pTchjet}{\ensuremath{p_{\mathrm{T}}^{\mathrm{ch\; jet}}}}
\newcommand{\pTfulljet}{\ensuremath{p_{\mathrm{T}}^{\mathrm{full\; jet}}}}
\newcommand{\pTtruth}{\ensuremath{p_{\mathrm{T,truth}}^{\mathrm{ch\; jet}}}}
\newcommand{\pTdet}{\ensuremath{p_{\mathrm{T,det}}^{\mathrm{ch\; jet}}}}
\newcommand{\Nevent}{\ensuremath{N_{\mathrm{event}}}}
\newcommand{\Ninc}{\ensuremath{N_{\mathrm{jets}}}}
\newcommand{\Sinc}{\ensuremath{\sigma_{\mathrm{jets}}}}

\title{Unravelling the Energy-Energy Correlators for Heavy Flavor Tagged Jets in \pp, \pPb~and \PbPb~ Collisions}

\date{\today  \hspace{1ex}}

\author{Ke-Ming Shen}
\affiliation{East China University of Technology, Nanchang 330013, China}
\affiliation{Key Laboratory of Quark $\&$ Lepton Physics (MOE) and Institute of Particle Physics, Central China Normal University, Wuhan 430079, China}
\author{Shi-Yong Chen}
\affiliation{Huanggang Normal University, Huanggang 438000, China}
\affiliation{Key Laboratory of Quark $\&$ Lepton Physics (MOE) and Institute of Particle Physics, Central China Normal University, Wuhan 430079, China}
\author{Yu-Jie Huang}
\affiliation{China University of Geosciences, Wuhan 430074, China}
\affiliation{East China University of Technology, Nanchang 330013, China}
\author{Wei Dai}
\email[Corresponding: ]{weidai@mail.cug.edu.cn}
\affiliation{China University of Geosciences, Wuhan 430074, China}
\author{Ben-Wei Zhang}
\affiliation{Key Laboratory of Quark $\&$ Lepton Physics (MOE) and Institute of Particle Physics, Central China Normal University, Wuhan 430079, China}

\begin{abstract}

In this study, energy-energy correlator (EEC) distributions for the $\rm D^0$, $\rm B^0$-tagged, inclusive, and the PYTHIA generated pure quark jets are computed in \pp, \pPb~and \PbPb~ collisions~at \sqrts $=5.02$~TeV for a same jet transverse momentum interval  $15<p_T^{\rm jet}<30$~GeV.  We find the number of particles per jet determines the height of the EEC distribution in \pp baseline. The averaged energy weight distribution resulted in a shift of the originally larger angular distributed particle distribution to a smaller \RL, thereby obtaining an EEC distribution. The EEC distributions for all quark-tagged jets in A+A exhibit a noticeable shift towards larger \RL~region, suggesting that the jets will be more widely distributed compared to those in \pp~ collisions. The jet quenching effect will cause the pair angular distribution to shift towards larger values and increase the number of particles per jet. This redistribution of energy within the jets suggests that the already reduced jet energy is redistributed among a larger number of particles, leading to a reduced energy weight per pair. The enhancement of the number of particles per jet and the reduced averaged energy weight interplay with each other to form the medium modification of EEC.

\end{abstract}


\maketitle

\section{Introduction}
\label{sec:Intro}

Jets originate from the fast partons generated during the initial hard processes at the Large Hadron Collider (LHC).  They suffer QCD radiation in a vacuum and split into a cluster of partons \cite{Appel:1985dq,Gyulassy:1990ye,Wang:1992qdg,JET:2013cls,Qin:2015srf}. 
It can be theoretically described by the perturbative QCD (pQCD) and modeled by Monte Carlo (MC) parton shower. 
Then they will be confined back into hadrons. 
The information of the final state hadrons inside the jets can be used to construct jet substructure observables~\cite{Chien:2015hda,Casalderrey-Solana:2016jvj,Kang:2016ehg,Tachibana:2017syd,KunnawalkamElayavalli:2017hxo,Luo:2018pto,Chang:2017gkt,Mehtar-Tani:2016aco,Milhano:2017nzm,Li:2017wwc,Li:2019dre,Wang:2019xey,Chen:2019gqo,Wang:2020bqz,Chen:2020tbl,Yan:2020zrz,Wang:2020ukj,Li:2021gjw,Luo:2021hoo,Zhang:2021sua,Chen:2022kic,Li:2022tcr,Zhang:2022bhq,Wang:2022yrp,Wang:2023eer,Zhang:2023jpe,Wang:2023udp,Kang:2023ycg,Zhou:2024khk,Wang:2024plm,Li:2024pfi}.
The measurements of the jet substructure in proton-nucleus (p+A) and nucleus-nucleus (A+A) collisions offer great opportunities to study the properties of nuclei and the quark-gluon plasma (QGP) through the interaction between the jets and the cold nuclear and hot QCD medium~\cite{ALICE:2019ykw,ALICE:2021njq,ALICE:2022vsz,ATLAS:2017zda,ATLAS:2022vii,CMS:2014jjt,CMS:2017qlm,CMS:2021iwu,LHCb:2019qoc,STAR:2020ejj,STAR:2021lvw,STAR:2021kjt}.

Benefit from its infrared-collinear (IRC) safe~\cite{Tkachov:1999py} and its calculability from first principles in QCD in the perturbative limit~\cite{Kardos:2018kqj}, 
the energy-energy correlators (EEC) offer precision probes of both perturbative and non-perturbative QCD dynamics in collisions ranging from $e^+e^-$ annihilation and hadronic collisions to deep inelastic scattering (DIS), as summarized in Ref.~\cite{Basham:1978bw,Larkoski:2013eya,Larkoski:2015uaa,Tulipant:2017ybb,Henn:2019gkr,Neill:2022lqx,Liu:2022wop,Andres:2023xwr,Kang:2023gvg}. 
In these studies, EEC observables can be defined in different ways.

In recent years, it has been extensively measured and studied for in \pp~ and A+A at the LHC energy \cite{ALICE:2019qyj,ALICE:2024dfl,CMS:2024ovv,Barata:2024bmx,Andres:2024pyz,Andres:2024hdd,Bossi:2024qho,Chen:2024cgx,Yang:2023dwc,Xing:2024yrb}.
They find its quark vs gluon separation power for jet quenching in inclusive charged jets \cite{Chen:2024cgx}, a wake signal found in $\gamma$-tagged jets \cite{Yang:2023dwc}, and the quark mass effect investigated in heavy flavor jets \cite{Xing:2024yrb}.

However, it is still lacking in unraveling studies on the observable itself, especially in heavy flavor tagged jets, the energy fraction possessed by the tagging heavy meson is quite dominant compared to the case in the light flavor category, making it fewer numbers of particle pairs inside the jets for heavy flavor tagged ones.  

The EEC observable arises from the angular scale, \RL. We also find it challenging to directly compare the energy-energy correlators (EECs) of different types of jets at the level of splitting observable, as observed in sub-jet measurements. These measurements can de-cluster jets down to basic splittings, such as Ref.~\cite{Zhang:2003wk,Djordjevic:2013pba,Craft:2022kdo,Dai:2022sjk,Zhang:2023oid,Andres:2023ymw,ALICE:2021aqk}, which directly reveal the Dead-Cone effect. It is worth investigating if the EEC observable has a similar discrimination power, especially down to the splitting level.

The remainder of this paper is organized as follows. 
In Sec.\ref{sec:pp} we test our \pp~ simulation setups against the ALICE data and then introduce our method to unravel the experimental definition of the EEC, compare the EEC distributions for $\rm D^0$, $\rm B^0$-tagged jets, the inclusive jets, and the pure quark jets on the particle pair level. We further explore the different contributions of unraveled factors to the EEC distribution. In Sec.\ref{sec:pA} we calculate and discuss the modification introduced by the implementation of the nPDFs in \pPb. We will include the hot QCD medium modifications in Sec.\ref{sec:AA} which is computed using the framework of the SHELL model~\cite{Dai:2018mhw}. The separated contributions of different unraveled factors are discussed as well. We will finally conclude in Sec.\ref{sec:Sum}.

\section{understanding the EEC observable in \pp}
\label{sec:pp}

In the ALICE experiment report \cite{ALICE:2024dfl}, the EEC is defined as an energy-weighted two-particle correlation as a function of the angular distance between particle pairs in a jet shown as in~Eq.~\ref{EEC-equ}.

\begin{widetext}
\begin{eqnarray}
\Sigma_{\text{EEC}}(R_{\text{L}}) = \frac{1}{ N_{\text{jet}}\cdot\Delta R}\int_{R_{\text{L}}-\frac{1}{2}\Delta R}^{R_{\text{L}}+\frac{1}{2}\Delta R}\sum_{\text{jets}}\sum_{i, j}\frac{p_{\text{T}, i}p_{\text{T}, j}}{p^2_{\rm T, jet}}\delta(R_{\text{L}}'-R_{\text{L}, ij})\dd R_{\text{L}}'~.
\label{EEC-equ}
\end{eqnarray}
\end{widetext}
where all final state particle pairs ($i$, $j$) inside each jet are summed up. The angular distance between each pair is defined in the $\eta-\varphi$ plane as $R_{\text{L}, ij} = \sqrt{(\varphi_j - \varphi_i)^2 + (\eta_j - \eta_i)^2}$, which $\Delta R$ is the angular bin width and $N_{\text{jet}}$ is the total number of jets.

In this work, we employ the Monte Carlo (MC) event generator PYTHIA 8~\cite{Sjostrand:2014zea} with Monash 2013 tune~\cite{Skands:2014pea} to simulate jet events in \pp~ collisions. To confront ALICE data at \sqrts$=13$~TeV~\cite{ALICE:2024node}, the final state charged hadrons are required to have $p_{T}>1$~GeV, as well as the $\rm D^0$-mesons and $\rm B^0$-mesons. Then the anti-$k_T$ algorithm from FASTJET package \cite{Putschke:2019yrg} is used to reconstruct jets with radius parameter $R=0.4$ for transverse momentum interval $15\sim 30$~GeV and the pseudorapidity interval is set to be $|\eta^{\rm jet}|<0.5$. $\rm D^0$, $\rm B^0$-tagged jets are further required to contain at least one $\rm D^0$-meson or $\rm B^0$-meson with $p_{\rm T}$ in the range of $5-30$~GeV.

\begin{figure}[!htb]
\centering
\includegraphics[width=9cm,height=11.5cm]{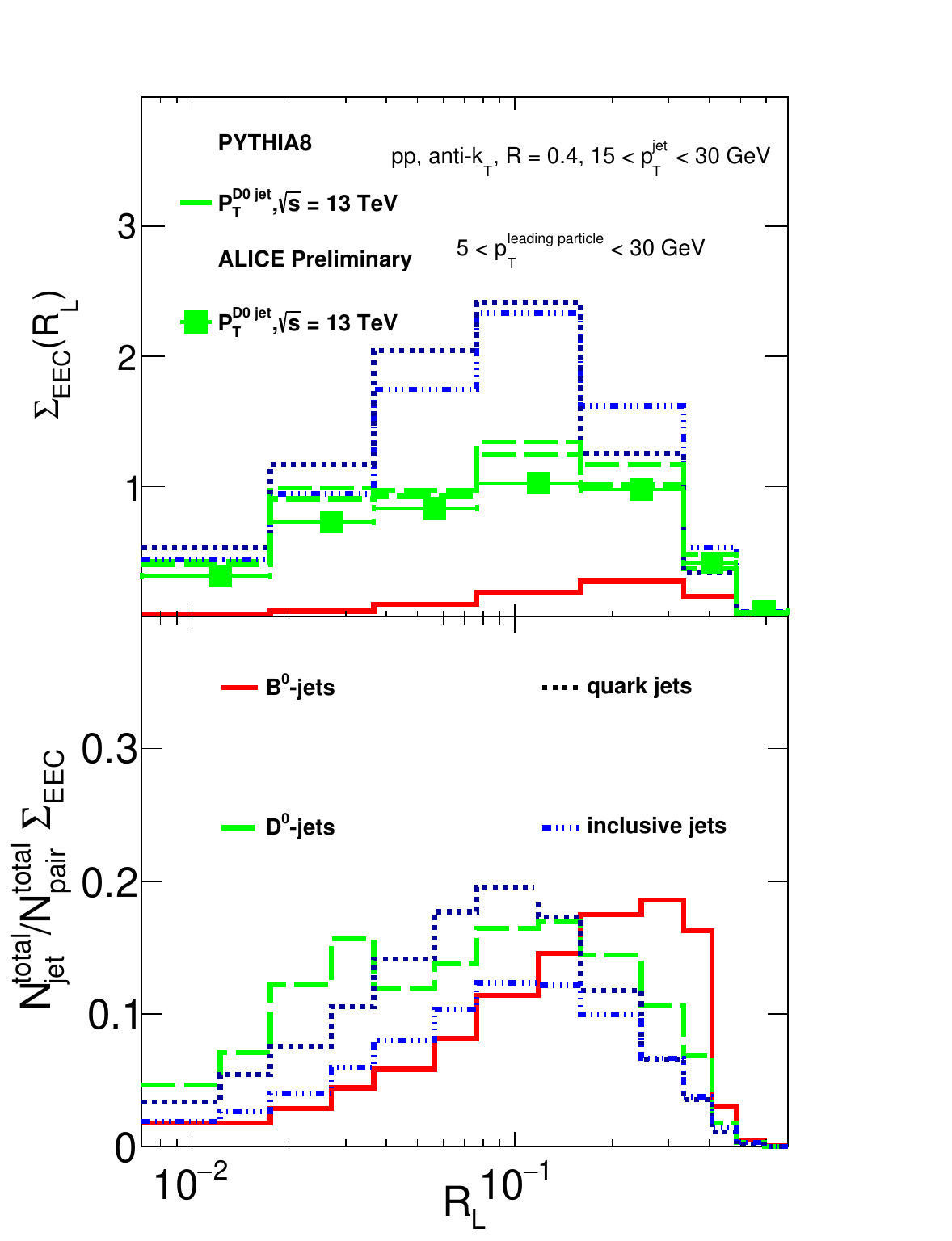}
\caption{Upper: number of jets normalized EEC as a function of $\RL$ for $\rm D^{0}$-meson tagged jets compared with ALICE data in \pp~ collisions at \sqrts$=13$~TeV \cite{ALICE:2024node} in the green line with dots. Results of $B^0$-tagged jets, $D^0$-tagged jets, light quark-initiated jets, and inclusive charged jets at \sqrts$=5.02$~TeVV are predicted.
Lower: the re-scaled EEC by a factor of $N^{\rm total}_{\rm pair}/N^{\rm total}_{\rm jet}$  based on the first factor in Eq.~\ref{EEC-fac} for 4 kinds of jets.}
\label{pp}
\end{figure}

\begin{table}[htb]
\caption{averaged number of particle pairs in jets: \pp}
\scalebox{1.2}[1.4]{
\begin{tabular}{c|c|c|c|c}
\hline
\hline
jets type &  $\rm B^0$- &  $\rm D^0$- & light quark- & inclusive \\
\hline
$\langle N_{\rm pair}^{\rm jet} \rangle_{\rm pp}$  & 1.3751 & 7.4908 &  13.2147  & 19.2787 \\
\hline
\hline
\end{tabular}
}
\label{table:ppNpair}
\end{table}

We compute and plot in the upper panel of Fig.~\ref{pp} EEC for $\rm D^0$-tagged jets at \sqrts $=13$ \GeV~to comfort with the experimental data.
Further, we compute at \sqrts $=5.02$~TeV the EEC for $\rm D^0$-, $\rm B^0$-tagged, inclusive, and the PYTHIA generated pure quark jets respectively in different line styles. We can immediately find a clear separation of the distributions even from the magnitude perspective. It seems that the heavier the tagging quark mass is, the lower the EEC distribution height is. However, such a comparison cannot be satisfied from the splitting angular point of view. At least, because of the tagging requirement and the different masses, we can find the difference of the particle numbers per jet for all these types of jets are quite distinct which is listed in Table.~\ref{table:ppNpair} and calculated by $N^{\rm total}_{\rm pair}/N^{\rm total}_{\rm jet}$. It has a serious impact on the magnitude of EEC distribution. For instance, the $\rm B^0$-tagged jets have much fewer particle numbers in a jet, therefore, the fewer number of particle pairs per jet will lead to a much lower EEC distribution, making the angular distribution comparison much harder. 
To demonstrate such a factor impacts the EEC observable and even to further unravel it into a splitting level, we try to rewrite the Eq.~\ref{EEC-equ} into Eq.~\ref{EEC-fac}
\begin{eqnarray}
\Sigma_{\text{EEC}}(R_{\text{L}}) = \frac{N^{\rm total}_{\rm pair}}{N^{\rm total}_{\rm jet}} \thinspace \cdot
\frac{\Delta N_{\rm pair}}{N^{\rm total}_{\rm pair} \Delta R}(R_{\rm L}) \thinspace \cdot \langle \rm weight \rangle(R_{L})
\label{EEC-fac}
\end{eqnarray}
where we average the energy weight term, $p_{\text{T}, i}p_{\text{T}, j}/p^2_{\rm T, jet}$, to every particle pair within every jet in each angular bin $\Delta R$, denoted as $\langle \rm weight \rangle$, therefore we can simply replace the integration and the sum over jets and $i, j$ into a summed number of pairs within each bin $\Delta R$, denoted as $\Delta N_{\rm pair}$. Consequently, Eq.~\ref{EEC-fac} appears as three factors: the averaged number of particles within a jet, normalized \RL~distribution of the number of pairs, and the averaged energy weight distribution as a function of \RL. It will benefit further discussion.

We immediately plot in the bottom panel of Fig.~\ref{pp} the EEC eliminating the impact of the different number of particles within a jet by rescaling the EEC by a factor of $N^{\rm total}_{\rm pair}/N^{\rm total}_{\rm jet}$ based on the first factor in Eq.~\ref{EEC-fac}, which is already calculated in Table.~\ref{table:ppNpair}. There is a caveat that we implement a minimum cut to the leading charged hadron $5<p_T^{\rm jet}<30$ GeV for the inclusive and pure light quark jet case for a fair comparison. We observed that the EEC distribution for $\rm B^0$-tagged jets peaks around \RL$=0.3$, that for $\rm D^0$-tagged jets peaks around \RL$=0.1-0.2$ and light quark jets case peaks around \RL$=0.09-0.1$, shown as a clear mass hierarchy. The broadening of the jet distribution increases with the mass of the tagging quark. We can certainly observe a clear dead-cone suppression at the lower \RL~region where $\rm B$ quark suppresses the most at a wider suppression region. However, we also caveat that the EEC distribution for $\rm D^0$-tagged jets at lower \RL~region will be complicated by the fact that we only require the jets containing at least one $\rm D^0$ meson which cannot exclude the contribution from gluon splitting process $g \to c \Bar{c}$. In such a process, $c \Bar{c}$ is produced in the same direction with smaller angular. 

We can also find that even though we have rescaled the EEC, there is still a magnitude difference for all 4 kinds of jets. 
This difference accumulated in each \RL~bin results in the fact that larger values of energy weight correspond to higher distributions, which is not normalized to unity at the same time.
The low distribution height of the inclusive jets indicates a finely divided energy fraction possessed by the particles inside the jets compared to that of heavy-flavor and pure light quark jets.

\begin{figure}[!htb]
\flushleft
\includegraphics[width=9.5cm,height=11cm]{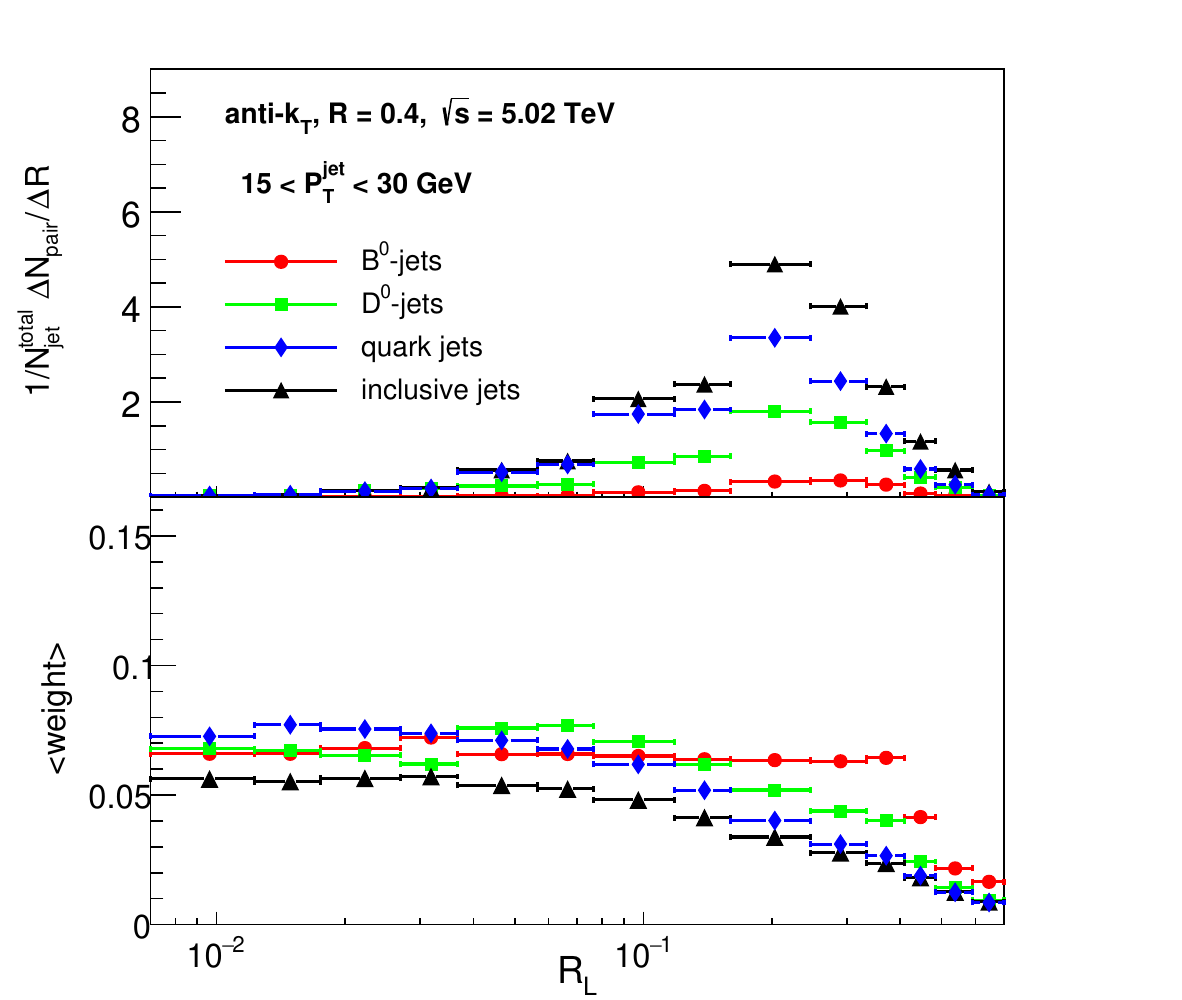}
\caption{The number of jets normalized angular distributions (upper panel), $\Delta N_{\rm pair}/(N_{\rm pair}^{\rm total}\Delta R)$, and the averaged energy weight factors (lower panel) for $\rm B^0$-tagged jets, $\rm D^0$-tagged jets, light quark-initiated jets, and inclusive charged jets are demonstrated for $15<p_T^{\rm jet}<30$ GeV in \pp collisions at \sqrts$=5.02$~TeV.}
\label{pp-ana}
\end{figure}

To demonstrate and compare the energy weight contribution to the EEC distribution of all 4 kinds of jets, the averaged energy weights in each \RL~bin are plotted in the bottom panel of Fig.~\ref{pp-ana} as a function of \RL. We find $\langle \rm weight \rangle$ of the $\rm D^0$, $\rm B^0$-tagged, and the light quark-initiated jets are similar even flatly distributed at smaller \RL~region where \RL$=0.007-0.1$ and then decreases with the increasing \RL~at larger \RL~region where \RL$=0.1-0.6$. In the larger region, more massive ones give a larger weight value, which indicates the jet energy is distributed to fewer particle pairs in more massive quark-tagged jets. The number of particle pairs in a jet definitely has an impact on it. Also interestingly, the $\langle \rm weight \rangle$ of $\rm B^0$-tagged jets remain flat till \RL$=0.4$ and then suddenly decrease rapidly with the increasing \RL. It reflects that the EEC distribution of the $\rm B^0$ meson involved pairs will reach its maximum at \RL$=0.4$ which matches the peak position displayed at the bottom panel of Fig.~~\ref{pp}. At \RL$ \geq 0.4$, it will involve pairs that do not contain $\rm B^0$ meson with a smaller $\langle \rm weight \rangle$ value.

We then take a look at the EEC distribution that does not accumulate energy weight but the pair numbers in \RL~bin in the upper panel of Fig.~\ref{pp-ana}, it also means we only take into account the first two factors in Eq.~\ref{EEC-fac}. We find the EEC distributions of the 4 types of jets peaks at a similar \RL~range. Combining the observation in the bottom panel of Fig.~\ref{pp-ana}, the contributions of the energy weight terms push the pair-angular distribution to a smaller angular value.
The lighter the tagging quark mass is, the smaller the angular peak will be pushed. We can summarize here that the energy weight term determines the distribution peak of EEC. The number of particle pairs in a jet determines the magnitude of the EEC distribution. We can also observe that a larger number of particle pairs per jet gives rise to higher pair-angular distribution height and lower energy weight values.

\section{nPDFs impact on the EEC observable in p+Pb}
\label{sec:pA}

Taking advantage of the discussion above, we calculate and discuss the nuclear parton distribution functions (nPDFs) impact on the EEC observable in p+Pb~collisions. It is not only to provide the baseline for the further discussion of hot QCD medium modification of EEC distribution in Pb+Pb~collisions but also to provide the baseline even in cold nuclear medium modification. In this study, we implement EPPS21 parameterization \cite{Eskola:2021nhw} to modify the parton distribution functions in the nucleus of Pb, therefore to calculate the EEC distributions of the $\rm D^0$, $\rm B^0$-tagged, inclusive, and the pure quark jets at the same time in p+Pb~and Pb+Pb~collisions. The other setups and the kinetic conditions are set to be the same as in the p+p~production. To benefit further presentation and discussion, we define the ratio of the EEC distribution as a function of \RL~in p+Pb~collisions and that in p+p~collisions as follows:
\begin{equation}
R_{\rm pA}^{\rm EEC}=\frac{\Sigma^{\rm pA}_{\rm EEC}(R_{\rm L})}{\Sigma^{\rm pp}_{\rm EEC}(R_{\rm L})}
\end{equation}

\begin{figure}[!htb]
\centering
\includegraphics[width=9cm,height=8cm]{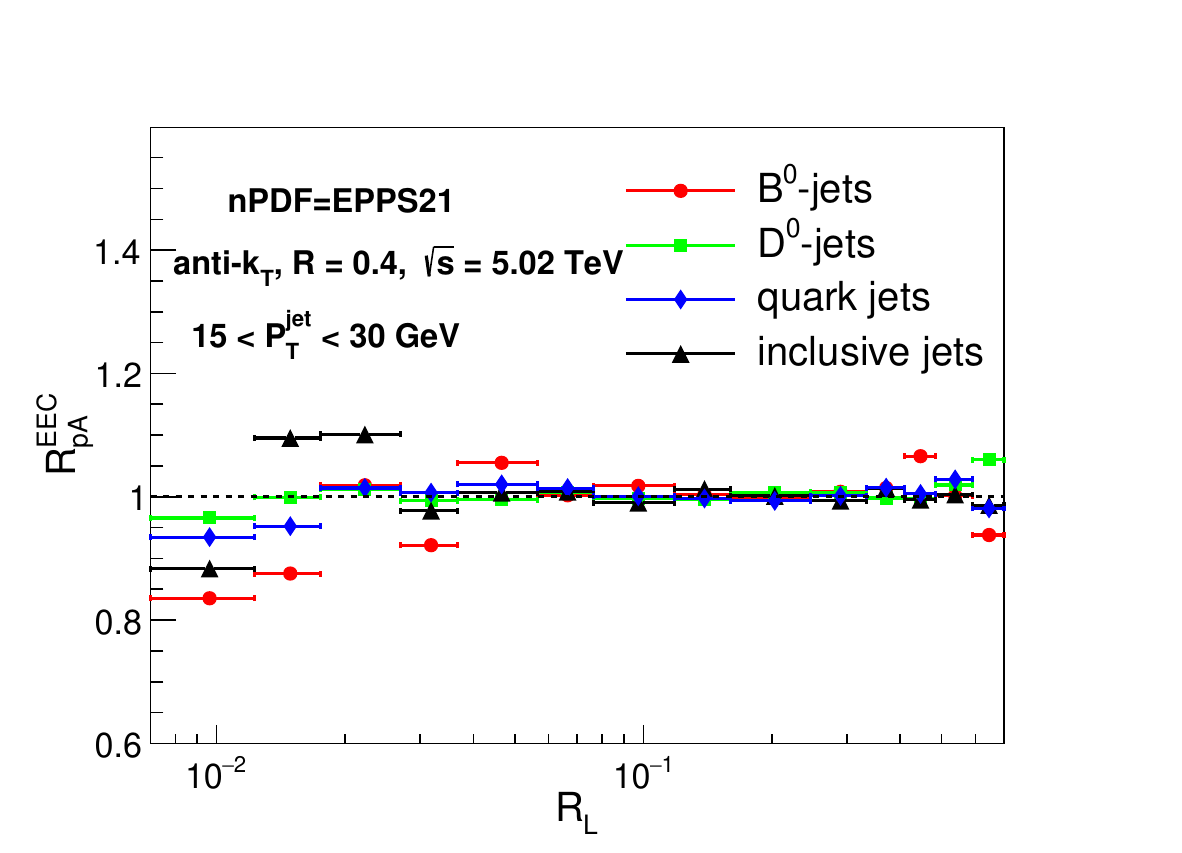}
\caption{The \pPb$/$\pp~ratios as functions of $R_{\rm L}$ for $\rm B^0$-tagged jets, $\rm D^0$-tagged jets, light quark-initiated jets and inclusive charged jets in \pPb~collisions at \sqrts$=5.02$~TeV using EPPS21 are plotted for the jet $p_{\rm T}$ intervals of jets: $15<p_{\rm T}^{\rm jet}<30$~GeV.
The line of unity is also plotted as a reference.}
\label{pA-all}
\end{figure}

We plot in Fig.~\ref{pA-all} the \pPb$/$\pp~ratios of the EEC distributions of $\rm B^0$-tagged jets, $\rm D^0$-tagged jets, quark jets, and inclusive charged jets as functions of \RL~at \sqrts$=5.02$~TeV~ with the jet \pT~interval of $15-30$~\GeV. We can observe that there can be clear suppression at the range of \RL$=0.007-0.02$ for all types of jets, especially for $\rm B^0$-tagged jets which gives the maximum suppression down to approximately $0.85$. In the range of \RL$=0.06-0.4$, the nPDFs-induced modification is nearly negligible. In the range of  \RL$=0.4-0.7$, we can not come to a confident conclusion.

\section{the jet quenching effect of the EEC observable in \PbPb}
\label{sec:AA}

\begin{figure*}[htb]
\centering
\includegraphics[width=18cm,height=6cm]{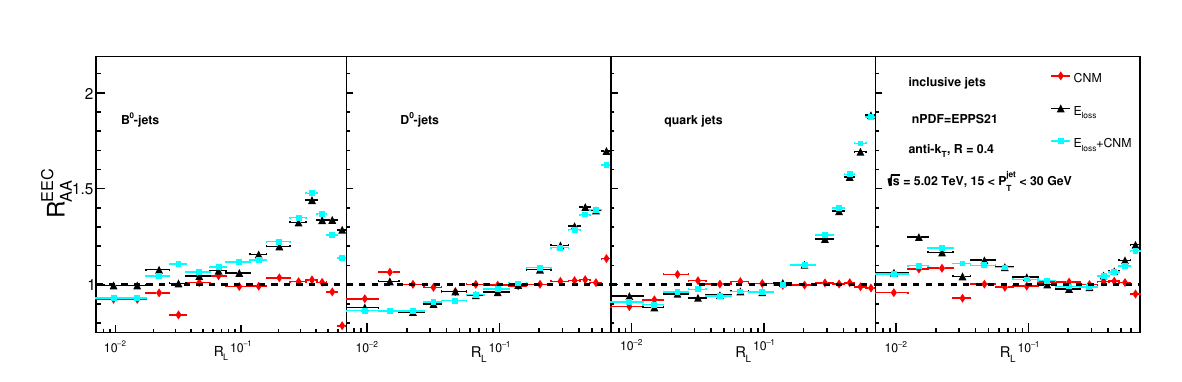}
\caption{The \PbPb/\pp~ratios of $R_{\rm AA}^{\rm EEC}$ as functions of $R_{\rm L}$ for $\rm B^0$-tagged jets, $\rm D^0$-tagged jets, light quark-initiated jets, and inclusive charged jets in the same $p_T^{\rm jet}$ regions (15$\sim$30~GeV) within the nPDFs of EPPS21 using PYTHIA 8.}
\label{AA}
\end{figure*}

In this section, we first introduce the theoretical model that is implemented for the simulation of the in-medium evolution of both light- and heavy-tagged jets. Since heavy quarks are viewed as effective hard probes to constrain the transport properties of QGP, A lot of theoretical models~\cite{Dainese:2004te,Lokhtin:2005px,Renk:2008pp,Zapp:2008gi,Armesto:2009fj,Schenke:2009gb,Zapp:2012ak,Wang:2013cia,Majumder:2013re,Xu:2014ica,Chesler:2014jva,Xu:2015bbz,Cao:2016gvr,Shi:2018izg} have been established to confront the experimental measurements, as seen in Refs.~\cite{Connors:2017ptx,Cao:2020wlm}.
In nucleus-nucleus collisions, the in-medium evolution of both light- and heavy-flavour partons are simulated by the SHELL model~\cite{Dai:2018mhw} in our work, which takes into account the elastic and inelastic partonic energy loss in the hot/dense QGP medium. 
The elastic scatterings between heavy quarks and thermal partons can be viewed as ``Brownian motion" in the limit of small momentum transfer~\cite{Xu:2015bbz}, which could be well described by the Langevin equations~\cite{Moore:2004tg,Cao:2013ita}. To take into account the inelastic interaction of heavy quarks during the propagation, we modified Langevin equations as follows:

\begin{eqnarray}
    \Delta \Vec{x}(t)=\frac{\Vec{p}(t)}{E}\Delta t, \\
    \Delta \Vec{p}(t)=-\eta_D\vec{p}\Delta t+\vec{\xi}(t)\Delta t-\vec{p}_g(t) ,
\label{LE}
\end{eqnarray}
Here $\eta_D$ is the drag coefficient used to control the energy dissipation strength of heavy quarks in the medium.
The stochastic term $\xi(t)$ denotes the random kicks as heavy quarks scatter with the medium partons, which obey a Gaussian distribution. These two terms describe the elastic interactions between the heavy quarks
and medium partons in the QGP.
In contrast, the last term of $-\vec{p}_g$ denotes the momentum correction caused by the medium-induced gluon radiation, which is sampled by the radiated gluon spectrum calculated within the Higher-Twist approach \cite{Majumder:2009ge,Guo:2000nz,Zhang:2003yn,Cao:2017hhk},
\begin{eqnarray}
    \frac{\dd N_g}{\dd x \dd k_\bot^2 \dd t}=\frac{2\alpha_s P(x)\hat{q}}{\pi k_\bot ^4}\sin^2\big(\frac{t-t_j}{2\tau_f}\big)\big(\frac{k_\bot^2}{k_\bot^2+x^2M^2}\big)^4 
\label{eq:g6}    
\end{eqnarray}
where $x$ and $k_\bot$ denote the energy fraction and transverse momentum of radiated gluon.
$P(x)$ is the QCD splitting function, which describes the splitting processes $g\to g+g$ and $q(Q)\to q(Q)+g$ \cite{Deng:2009ncl}.
$\tau_f=2Ex(1-x)/(k_\bot^2+x^2M^2)$ stands for the formation time of the daughter gluon.
$\hat{q}$ is the general jet transport parameter in the QGP \cite{Chen:2010te}.
The last term describes the suppression factor results from the ``dead-cone" effect of heavy quarks with mass M \cite{Dokshitzer:2001zm,ALICE:2019nuy}.

 The gluon radiation probabilities in QGP of both quarks and gluons  during a given Langevin evolution time interval ($\Delta t$) are calculated from the Poisson probability distribution: 
\begin{eqnarray}
P_{rad}(t,\Delta t)=1-e^{-\left\langle N(t,\Delta t)\right\rangle} \, .
\label{eq:g7}
\end{eqnarray}
where $\left\langle N(t,\Delta t)\right\rangle$ is the averaged number of radiated gluons and can be calculated by integrating the gluon radiation spectrum shown in Eq.~\ref{eq:g6}. If radiation happens, the number of radiated gluons is sampled from a Poisson
distribution:
 \begin{eqnarray}
P(n_{g},t,\Delta t)=\frac{\left\langle N(t,\Delta t)\right\rangle^{n_{g}}}{n_{g}!}e^{-\left\langle N(t,\Delta t)\right\rangle} \, .
\label{eq:g8}
\end{eqnarray}
For massless partons, the collisional energy loss is estimated by the pQCD calculations within the Hard-Thermal Loop approximation \cite{Neufeld:2010xi}: $\dd E^{\rm coll}/\dd t=\frac{\alpha_{s}C_{s}\mu^{2}_{D}}{2}\ln\frac{\sqrt{ET}}{\mu_{D}}$, where $\mu_{D}$ is the Debye screening mass.

The hydrodynamics time-space evolution of the QGP medium is described by the CLVisc programs \cite{Shen:2014vra}, which provides the temperature and velocity of the expanding hot and dense nuclear matter. All the showered partons will stop their propagation in the QGP medium when the local temperature falls below $T_c=165$~MeV. The colorless method developed by the JETSCAPE collaboration\cite{Putschke:2019yrg}
is used to construct strings from all the final state partons, then we
perform hadronization and hadron decays using the PYTHIA Lund string method.
The SHELL model could provide satisfactory descriptions of a series of jet observables in high-energy nuclear collisions, such as $p_T$ imbalance~\cite{Dai:2018mhw}, correlations of $Z^0$+HF hadron/jet~\cite{Wang:2020qwe}, radial profiles~\cite{Wang:2019xey,Wang:2020ukj} and jet angularities~\cite{Yan:2020zrz,Chen:2022kic,Wang:2024plm,Li:2024pfi}.

We then define the \PbPb$/$\pp~ratio of the EEC distributions to demonstrate the jet quenching effect on the EEC observable, shown as follows:
\begin{equation}
R_{\rm AA}^{\rm EEC}=\frac{\Sigma^{\rm AA}_{\rm EEC}(R_{\rm L})}{\Sigma^{\rm pp}_{\rm EEC}(R_{\rm L})}
\end{equation}
We plot in the 4 panels of Fig.~\ref{AA} the $R_{\rm AA}^{\rm EEC}$ of the $\rm B^0$-tagged jets, $\rm D^0$-tagged jets, quark-initiated jets, and inclusive charged jets as functions of \RL~in \PbPb~collisions at \sqrts$=5.02$~TeV in the $p_T^{\rm jet}$ region (15$\sim$30~GeV). We present the results solely considering the implementation of nPDFs, exclusively incorporating the hot QCD medium modification as well as the combination of the two. 

We start the analysis by first comparing the impact of nPDFs on the EEC of 4 types of jets in A+A collisions denoted by lines with diamond points in the figure and find the overall modification is mild, with slight suppression in the \RL~region of $0.007-0.03$ and smaller turbulent at larger \RL~region of $0.5-0.7$. One can observe a clear suppression at the smaller \RL~region of $0.007-0.04$ and larger \RL~region of $0.5-0.7$ for $\rm B^0$-tagged jets, which will consequently affect the total $R_{\rm AA}^{\rm EEC}$ at this region. 

Then let us focus on the hot QCD medium modification of the EEC. By comparing the results plotted by the lines with square points in the 4 panels, we observe that the EEC distributions shift to larger \RL regions, indicating that jets are distributed more broadly in A+A collisions compared to \pp collisions. Since we have already carefully discussed in the previous work \cite{Chen:2024cgx} that inclusive charged jets are dominated by gluon-initialed jets which suffer a signature enhancement at smaller \RL~region (unlike quark-initiated jets) as we can see in the very right panel of Fig.~\ref{AA}, we will mainly focus on the other three panels to discuss the medium modification of the EEC for heavy flavor jets with the pure quark-initiated jets as reference. For the EECs of all three quark jets, they all suffer suppression at smaller \RL~region and enhancement at larger \RL~region. 

Talking about enhancement, quark-initiated jets exhibit the strongest amplification in the \RL~region of $0.15-0.7$. Following closely are $\rm D^0$-tagged jets, with a higher \RL~value corresponding to a greater level of enhancement. Interestingly, the EEC for $\rm B^0$-tagged jets suffer less enhancement with a wider \RL~region of $0.03-0.7$. Unlike the other two lighter quark counterparts, its enhancement will reach a maximum at \RL$=0.3-0.4$ and begin to fall with further increasing \RL. Then we look at the suppressions for the 3 types of jets. The findings indicate the absence of a mass-hierarchical pattern in terms of the extent of suppression. The EEC for $\rm D^0$-tagged jets is suppressed the most but the suppression for $\rm B^0$-tagged jets is mild. 

\begin{figure}[!htb]
\centering
\includegraphics[width=9cm,height=10.5cm]{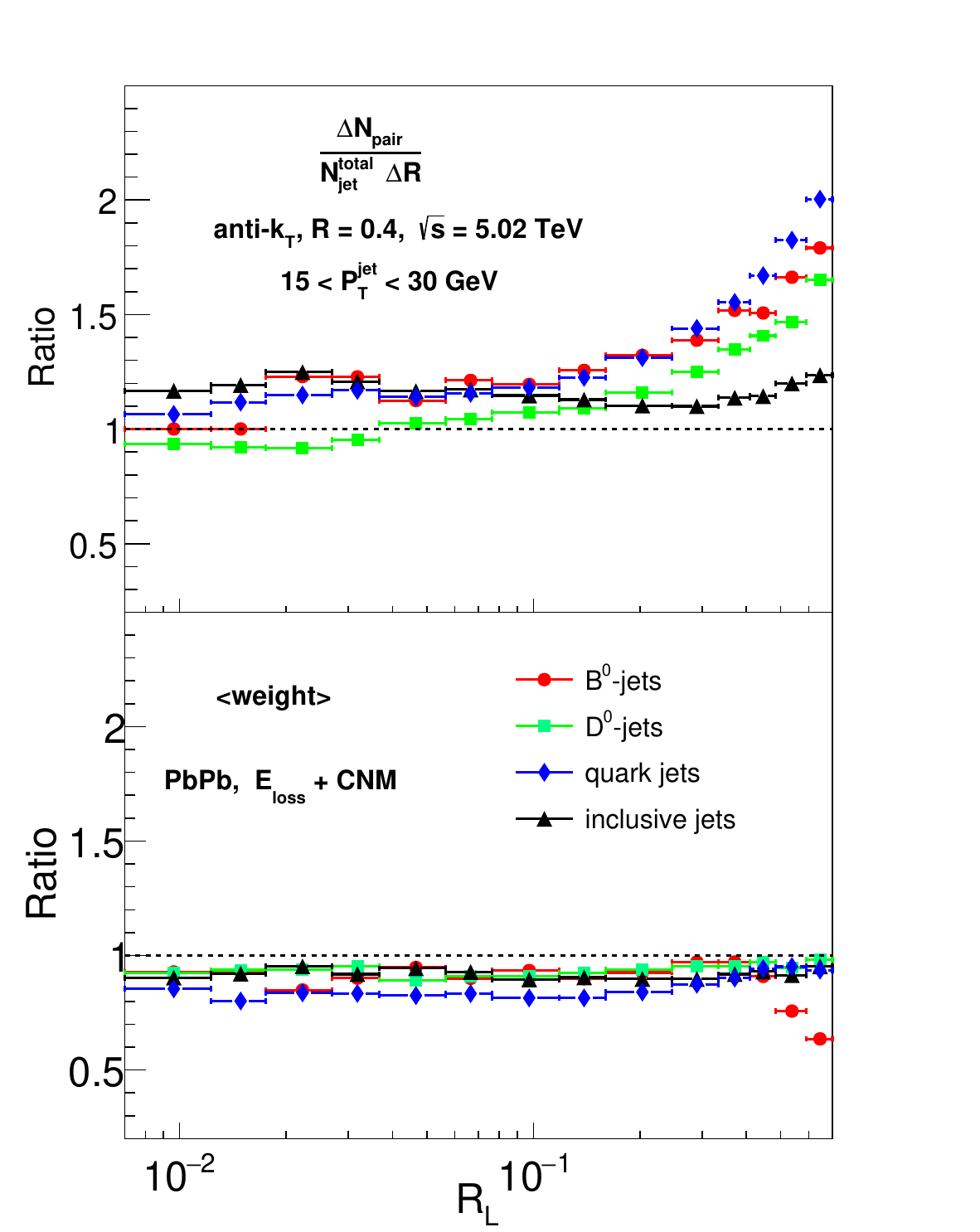}
\caption{ The \PbPb/\pp~ratios of the jet number normalized angular distributions $\Delta N_{\rm pair}/(N_{\rm pair}^{\rm total}\Delta R)$ (upper panel), and \PbPb/\pp~ratios of the averaged energy weight factors (lower panel) for $\rm B^0$-tagged jets, $\rm D^0$-tagged jets, light quark-initiated jets, and inclusive charged jets are demonstrated for $15<p_T^{\rm jet}<30$ GeV in \PbPb~collisions at \sqrts$=5.02$~TeV, respectively.}
\label{AA-ana}
\end{figure}

The simultaneous consideration of enhancement and suppression reveals a counter-intuitive finding: the enhanced distribution at larger angles is not fully restored at smaller angles to suppression. We realize in Sec.~\ref{sec:pp} that the EEC observable is not normalized to itself and also it is the weight that filled into each \RL~bin. Therefore, the pair angular distribution, the averaged number of pairs per jet, and the averaged energy weight all impact the jet quenching of EECs. We then in Fig.~\ref{AA-ana} plot the \PbPb$/$\pp~ratios of the number of jets normalized pair-angular distribution in the left panel and the \PbPb$/$\pp~ratios of the averaged energy weight in the right panel. We also list the value of the averaged number of particle pairs per jet for all 4 types of jets produced in the \PbPb~collisions in Table.~\ref{table-Npair}, their values of the \PbPb$/$\pp~ratios are listed in the next row.

\begin{table}[htb]
\caption{Averaged number of particle pairs in jets: \PbPb}
\scalebox{1.2}[1.2]{
\begin{tabular}{c|c|c|c|c}
\hline
\hline
jet type &  $\rm B^0$- &  $\rm D^0$- &  light quark-  & inclusive \\
\hline
$\langle N_{\rm pair}^{\rm jet} \rangle_{\rm PbPb}$  & 1.8902 & 8.9286 &  17.6994  & 21.7357 \\
\hline
AA/pp & 1.3746 & 1.1919 & 1.3393 & 1.1274 \\
\hline
\hline
\end{tabular}
}
\label{table-Npair}
\end{table}

We start with the averaged number of particle pairs per jet in Table.~\ref{table-Npair}, the first row is the same as in Table.~\ref{table:ppNpair} denoted as $\langle N_{\rm pair}^{\rm jet} \rangle_{\rm PbPb}$, the averaged number of particle pairs per jet for $\rm B^0$-tagged jets is still less than $2$, the value for the jets with lighter tagging quark is becoming larger. However, it is the \PbPb$/$\pp~ratio that can help us understand the medium modification to EEC. We find in the second row that the medium modification will always enhance the final state particle numbers in jets compared to that in \pp. The value of ratios for $\rm B^0$-tagged and light quark-initiated jets are similar, around $1.33-1.38$, and the value of ratios for $\rm D^0$-tagged and inclusive charged jets are similar, around $1.12-1.2$. There seems no mass hierarchy in this part.

In the upper panel, we compare the \PbPb$/$\pp~ratios of the number of jets normalized pair-angular distributions for 3 types of quark jets and neglect discussion of the plot for inclusive jets since it has been investigated in Ref.~\cite{Chen:2024cgx}. We find the enhancement of the EEC for the light quark-initiated jets is the largest and the enhancement for $\rm B^0$-tagged
jets is the second, the enhancement for $\rm D^0$-tagged jets is much weaker. The $\langle N_{\rm pair}^{\rm jet} \rangle_{\rm PbPb}$ certainly will have its contribution in this. Since the number of pairs normalized pair-angular distributions are the ones normalized to unity, It is the $\langle N_{\rm pair}^{\rm jet} \rangle_{\rm PbPb}$ explains the overall \PbPb$/$\pp~ratios of the number of jets normalized pair-angular distributions to be larger than unity. This set of plots shows a clear and unified broad shift of the pair-angular distributions towards larger \RL. 

We then compare \PbPb$/$\pp~ratios of the average energy weight for all 4 types of jets. A relatively flat distribution of the modification has been observed, and all the curves are under unity, reflecting not only the loss of the total energy of the jet but also the increasing number of particles in the jet. It corresponds exactly to the Table.~\ref{table-Npair}, the larger the value of the \PbPb$/$\pp~ratio for the $\langle N_{\rm pair}^{\rm jet} \rangle_{\rm PbPb}$ is the lower the suppression of the average energy weight is. We take $\rm D^0$-tagged jets for instance and find that the \PbPb$/$\pp~ratio of the $\langle N_{\rm pair}^{\rm jet} \rangle_{\rm PbPb}$ is relatively smaller, the \PbPb$/$\pp~ratios in this figure is closer to unity. We also find the interesting part of $\rm B^0$-tagged
jets, the \PbPb$/$\pp~ratio for it dramatically drop with the increasing \RL~when the value of \RL~is larger than $0.4$. It explains the feature of the maximum enhancement approximately at \RL$=0.4$ in Fig.~\ref{AA}.

\section{Summary}
\label{sec:Sum}

In this work, we unravel the experimental definition of the EEC observable into three factors: the number of particles within a jet, the number of pairs normalized \RL~distribution, and the averaged energy weight distribution as a function of \RL. EEC distributions for the $\rm D^0$, $\rm B^0$-tagged, inclusive, and the PYTHIA generated pure quark jets are computed in \pp, \pPb~ and \PbPb~at \sqrts $=5.02$~TeV for a same jet transverse momentum interval  $15<p_T^{\rm jet}<30$~GeV. We find the $N^{\rm total}_{\rm jet}/N^{\rm total}_{\rm pair}$ rescaled EEC can be used to demonstrate their mass hierarchy in the \pp~ baseline. The number of particles per jet determines the height of the EEC distribution. The averaged energy weight distribution decreases with the increasing \RL. It helps the originally larger angular distributed particle distribution shift to smaller \RL~to obtain an EEC distribution. 

The subsequent analysis investigates the impact of implementing nPDFs in \pPb~ collisions. Our findings reveal a consistent suppression within the range of \RL$=0.007-0.02$ across all types of jets. Notably, for jets tagged with $\rm B^0$ meson, $R_{\rm pPb}$ exhibits the most significant level of suppression, reaching 0.85. The modification is either insignificant or turbulent at the opposite end of the range in \RL.

The EEC distributions for all quark-tagged jets in A+A exhibit a noticeable shift towards a larger \RL~region, suggesting that the jets will be more widely distributed compared to those in \pp. The light quark-initiated jets exhibit the strongest enhancement in the \RL~region of $0.15-0.7$. Following closely are $\rm D^0$-tagged jets, with a higher \RL~value corresponding to a greater level of enhancement. The EEC for $\rm B^0$-tagged jets suffer less enhancement with a wider \RL~region of $0.03-0.7$. The enhancement of this quark will reach a maximum at \RL$=0.3-0.4$ and subsequently decline as \RL\ increases further. The presence of a mass-hierarchical pattern in terms of the degree of suppression is not observed.

The jet quenching effect on the EEC observable was analyzed based on three unraveled factors. The results indicate a shift in the pair angular distribution to larger \RL~values and an increase in the number of particles per jet. This redistribution of energy within the jets suggests that the already energy-lost jet energy is distributed among a larger number of particles, leading to a reduced energy weight per pair. The enhancement of the number of particles per jet and the reduced averaged energy weight interplay with each other. The \PbPb$/$\pp~ratio of the averaged energy weight for $\rm B^0$-tagged jets dramatically drop with the increasing \RL~when the value of \RL~increases beyond $0.4$. It explains the characteristic maximum enhancement approximately at \RL$=0.4$ in $R_{\rm AA}^{\rm EEC}$.

{\bf Acknowledgments:}  This work was supported by the Doctoral Research of ECUT (Nos. DHBK2019211).
This research is also supported by the Guangdong Major Project of Basic and Applied Basic Research No. 2020B0301030008, and the National Natural Science Foundation of China with Project Nos. 11935007 and 12035007.

\vspace*{-.6cm}

%


\end{document}